# Model-aided quantification of patient-specific benefit in mitigating radiation induced lymphopenia by particle therapy of cancer

Vladislav Sandul, M.Sc.[*,†], Marco Durante, PhD[*,†], Thomas Friedrich, PhD[*]

[*] Biophysics Department, GSI Helmholtzzentrum für Schwerionenforschung GmbH, Darmstadt, Germany

[†] Institute for Condensed Matter Physics, Technical University Darmstadt, Darmstadt, Germany

**Corresponding Author:** Thomas Friedrich, t.friedrich@gsi.de

**Author Responsible for Statistical Analysis:** Vladislav Sandul, v.sandul@gsi.de

**Conflict of Interest:** None

**Funding Statement:** This work was supported by Helmholtz Graduate School for Hadron and Ion Research (HGS-HIRe) for FAIR.

**Data Availability Statement:** Research data are available at https://www.gsi.de/bio-alc

# Model-aided quantification of patient-specific benefit in mitigating radiation induced lymphopenia by particle therapy of cancer

## Abstract

Treatment-related lymphopenia is a frequent and clinically significant consequence of cancer therapy that can compromise immune-mediated tumor control and worsen patient outcomes. Despite its importance, no mechanistic framework exists to accurately predict the severity of lymphopenia from patient-specific data. Here, we present a biokinetic model that quantitatively describes lymphocyte depletion and recovery during and after radiotherapy, integrating radiation dose–volume distributions, blood circulation dynamics, and distinct kinetics of fast- and slow-recovering lymphocyte populations. The model was calibrated and validated using 56 independent clinical datasets encompassing various tumor sites and treatment modalities. It reproduces observed lymphocyte counts and enables prediction of individual severity of lymphopenia from baseline or early-treatment counts. Applying this framework, we demonstrate that particle therapy reduces lymphocyte depletion by ~30% compared with photon therapy, providing a quantitative explanation for its observed immune-sparing benefit. By linking radiation physics, immune kinetics, and clinical outcomes, our model establishes a mechanistically grounded predictive approach for anticipating systemic immune toxicity. Beyond radiotherapy, this framework offers a generalizable strategy for integrating early biological markers into treatment optimization, advancing personalized and immune-preserving cancer therapy.

## Introduction

Radiation-induced lymphopenia (RIL) is a frequently observed side effect of radiation therapy, characterized by a marked reduction in blood lymphocyte counts [1–3]. As RIL correlates with poor overall survival (OS) and increased disease recurrence rates [2,4–8], it has clinical relevance, although this correlation is not necessarily causative. In the context of modern combined therapies employing immunotherapy in combination with radiation, RIL can compromise immunotherapy efficacy, which relies on a functional immune system for antitumor responses [9].

Recent investigations have identified several factors contributing to RIL development. These include baseline absolute lymphocyte count (ALC), planning target volume (PTV), mean body dose, and radiation exposure to blood, bone marrow, spleen, lymph nodes, and various organs including the heart, lungs, and liver [3,7,10–16]. Treatment modality also influences RIL severity, with proton and ion therapy demonstrating superior lymphocyte preservation compared to conventional photon therapy [17–23]. Cancer type and location have emerged as another significant factor [16]. Despite extensive research on this topic during the last decade, a conclusive understanding of radiation-induced lymphocyte depletion remains to be established.

Mathematical models have been widely applied to studies of RIL. A comprehensive review of existing models can be found in Cella et al. [24], with additional insights from Paganetti [10]. Early biokinetic models estimated blood and bone marrow dose distribution during extracorporeal irradiation and lymphocyte turnover between blood and lymphatic compartments [25,26]. Subsequent developments incorporated detailed trafficking patterns and multiple compartments [27–35]. However, many of these models rely on a large number of parameters, which may lead to overfitting, limited predictivity, biologically unrealistic parameter values, or an inability to reproduce key features of the observed ALC dynamics.

Many studies on RIL have shown that ALC follows an exponential-like decrease in the blood during radiotherapy, typically not dropping to zero, with ALC level at the end of treatment being ~20%-30% of the initial level [16]. After therapy, ALC tends to increase rapidly over the course of several months before recovery significantly slows down. Subsequent recovery progresses at a markedly reduced rate, as supported by long-term observations [36,37], indicating that full recovery may take several years. To explain ALC dynamics during and after radiotherapy, we proposed in [16] two hypotheses of radiation interaction with the lymphatic and lymphopoietic system: (1) Damage to the lymphopoietic system, affecting overall lymphocyte production after therapy and (2) lymphocyte classes with different recovery rates. In this study, we follow the second hypothesis, since the feasibility of validating lymphopoietic damage from ALC data is limited.

We introduce a biokinetic model describing lymphocyte count dynamics during and after radiotherapy, incorporating two lymphocyte classes: fast-recovering (FRL) and slow-recovering (SRL). The model accounts for both radiation-induced and natural lymphocyte depletion, as well as lymphocyte replenishment. A distinctive feature of the proposed model is its relative simplicity and small parameter number (only one after fixing physiological parameters), designed to minimize assumptions while still enabling quantitative, interpretable results. Despite its simplicity, the model successfully reproduces key features of lymphocyte dynamics during and after radiotherapy, including an exponential-like decline to a non-zero plateau in the period of treatment, followed by a biphasic post-treatment recovery. The model was benchmarked on GSI ALC database - a large dataset of ALC measurements during and after radiotherapy [16,38].

The model does not only reproduce observed ALC dynamics but also enables personalized prospective prediction of lymphopenia severity, based on clinical experience. Defining the model's parameters on retrospective clinical data, it predicts the expected RIL grade from pre- or early-treatment ALC and quantifies the benefit of particle therapy over photons in mitigating immune depletion. This allows estimates of how many patients would benefit and individual predictions to support personalized treatment modality selection.

## Materials and Methods

### *Biokinetic RIL model*

Assuming rapid exchange of lymphocytes between the blood and the lymphatic system, we consider the redistribution of lymphocytes across compartments to establish a dynamic equilibrium within the typical 1-day interval between radiotherapy fractions. Consequently, the model describes the temporal dynamics of lymphocytes in the entire body, without focusing on individual compartments. The lymphocyte population is divided into two classes – fast-recovering lymphocytes (FRL) and slow-recovering lymphocytes (SRL) – and therefore the overall lymphocyte dynamics $L(t)$ is expressed as

$$L(t) = L_{\mathrm{S}}(t) + L_{\mathrm{F}}(t), \quad (1)$$

where $L_{\mathrm{S}}(t)$ and $L_{\mathrm{F}}(t)$ represent the SRL and FRL dynamics, respectively.

The ALC dynamics in the model is based on the assumption that the temporal change in the lymphocyte count results from competition between the constant production rate of new cells and the death rate of existing cells arising from either natural causes or radiation exposure:

$$\frac{dL_i(t)}{dt} = -(k_{\mathrm{Rad},i}(t) + k_{\mathrm{Nat},i})L_i(t) + k_{\mathrm{Nat},i}\, L_{i,0} \qquad (2)$$

Here $L_i(t)$ represents the number of lymphocytes of type $i$ in the blood at time $t$, while $L_{0,i}$ denotes the pre-treatment level of this lymphocyte type. The term $k_{\mathrm{Rad},i}(t)$ is the radiation-induced depletion rate, modeled as a constant $k_{\mathrm{Rad},i}$ during the radiotherapy period $0 < t < T$ with treatment duration $T$, and zero afterwards. The parameter $k_{\mathrm{Nat},i}$ corresponds to the natural depletion rate of lymphocytes of type $i$, and its inverse $1/k_{\mathrm{Nat},i}$ reflects the average lifespan of these lymphocytes. Notably, the production rate of lymphocytes is a physiological constant, described under equilibrium conditions before irradiation as $k_{\mathrm{Nat},i}\, L_{i,0}$.

The solution of (2) is

$$L_i(t) = \begin{cases} L_{0,i}\left(\frac{k_{\mathrm{Rad},i}}{k_{\mathrm{Nat},i} + k_{\mathrm{Rad},i}} e^{-(k_{\mathrm{Nat},i} + k_{\mathrm{Rad},i})t} + \frac{k_{\mathrm{Nat},i}}{k_{\mathrm{Nat},i} + k_{\mathrm{Rad},i}}\right), \ 0 \le t \le T \\ L_{0,i} - \left(L_{0,i} - L_{\mathrm{EoT},i}\right)e^{-k_{\mathrm{Nat},i}(t-T)}, \ t > T \end{cases} \quad (3)$$

where $L_{\mathrm{EoT},i} = L_i(t = T)$ is the amount of lymphocytes of type $i$ at the end of RT.

Since $L(t)$ is described in arbitrary units, normalization can be applied by setting $L(0) = L_0 = 1$. Introducing the parameter $\gamma$ representing the fraction of FRL within the total lymphocyte population before therapy, $\gamma = L_{0,\mathrm{F}}/L_0$, the initial conditions are defined as $L_0 = 1, L_{0,\mathrm{F}} = \gamma L_0, L_{0,\mathrm{S}} = (1-\gamma)L_0$. In this study we assume that both subgroups of lymphocytes, SRL and FRL, exhibit the same radiosensitivity, therefore, $k_{\mathrm{Rad,S}} = k_{\mathrm{Rad,F}} \equiv k_{\mathrm{Rad}}$.

With these assumptions and given that the therapy duration $T$ is typically known, the total number of free parameters in the model is reduced to three physiological parameters $\gamma$, $k_{\mathrm{Nat,S}}$ and $k_{\mathrm{Nat,F}}$ and one case-specific parameter $k_{\mathrm{Rad}}$. This parameterization simplifies the model while maintaining sufficient flexibility to describe lymphocyte dynamics during and after radiotherapy.

*Data filtering to benchmark model*

We utilized the GSI ALC database, comprising ALC curves obtained from published studies on RIL [38]. To ensure the quality and comparability of the datasets selected for model fitting, a systematic filtering process was applied with the following criteria: (i) Individual patient data were excluded due to significant fluctuations in measured ALC values, which could obscure general trends and prevent reliable parameter estimation.

(ii) To avoid redundancy and information duplication, datasets aggregating all patients within a study were used for further analysis, and available subsets are discarded if they are stratified by outcome-related clinical parameters such as post-treatment OS, RIL grade, cancer stage, baseline ALC, or unrelated comorbidities. Conversely, subsets categorized specifically by radiation modality or cancer type were kept for analysis to maintain specificity, while corresponding aggregated datasets were discarded. (iii) Datasets with ALC as a function of accumulated dose rather than time were excluded.

Additionally the study of Mohan et al. [39] was excluded due to its unique ALC increase after the start of radiotherapy, which deviates significantly from typical trends observed in other studies. The dataset of Schad et al. [37] was not included since no data on the lymphocyte nadir or ALC during radiotherapy are reported. The dataset on TBI patients from Ellsworth et al. [40] was not used because of extreme rate of lymphocyte depletion, but is analyzed separately to investigate lymphocytes radiosensitivity in more detail. The summary of the included / excluded datasets and corresponding publications is given in table S1 in the supplementary material.

Following these criteria, 56 datasets from patient cohorts remained for model benchmarking and validation. The selected datasets were normalized to initial lymphocyte count $L_0 = 1$. This reduced model complexity by decreasing the number of free parameters and enhanced the accuracy and interpretability of the remaining fitted model parameters.

*Fitting procedure and parameters estimation*

The fitting procedure was conducted using non-linear regression with the Levenberg-Marquardt method in *Wolfram Mathematica*, organized into two consecutive steps to separate the estimation of physiological parameters from case-specific radiation-induced depletion rates. To enhance stability and robustness of the fitting process, both the ALC data and the model were logarithmically transformed before fitting, so that the uncertainties of ALC measurements are expected to be more uniform, given that the underlying dynamics are approximately exponential.

In the first step, fits were performed on a subgroup of filtered datasets containing at least six data points, including at least one data point during treatment (in addition to the fixed baseline RLC point $L_0 = 1$) and at least two data points collected after the end of therapy. This subgroup comprised 23 datasets, and the model was fitted separately to each of them. The constraints $0 \leq \gamma \leq 1$, $0.1 \leq k_{\mathrm{Nat,F}} \leq 5\ \mathrm{wk}^{-1}$, $0 \leq k_{\mathrm{Nat,S}} \leq 0.1\ \mathrm{wk}^{-1}$, and $0\ \mathrm{wk}^{-1} \leq k_{\mathrm{Rad}}$ were applied to ensure biological plausibility, as they reflect physiological conditions: The constraint on $\gamma$ arises from its definition as the fractional contribution of FRL. Negative depletion or recovery rates are not meaningful. The upper boundary for $k_{\mathrm{Nat,S}}$ of 0.1 $\mathrm{wk}^{-1}$ was introduced to distinguish SRL from FRL, and the limit of $k_{\mathrm{Nat,F}} \leq 5$ $\mathrm{wk}^{-1}$ excludes unrealistically rapid recovery corresponding to complete turnover in less than 1.5 days.

Weighted mean values of the three physiological parameters, $\gamma$, $k_{\mathrm{Nat,F}}$, and $k_{\mathrm{Nat,S}}$, were then computed, with weights inversely proportional to squared fitting uncertainties. Parameter uncertainties were derived from the covariance matrix of the fit, or, if unrealistically small by coincidence, replaced by the mean uncertainty across all experiments (see Fig. S1 in the supplementary material). Histograms of the fitted parameters are shown in Fig. 1(a–c). The weighted mean values of the parameters were found to be $\gamma = 0.39 \pm 0.08$, $k_{\mathrm{Nat,F}} = 0.22 \pm 0.20\ \mathrm{wk}^{-1}$ and $k_{\mathrm{Nat,S}} = (3.4 \pm 2.5) \cdot 10^{-3}\ \mathrm{wk}^{-1}$. Individual parameter values for each dataset and corresponding uncertainties are provided in Fig. S2 of the supplementary material.

By converting the natural depletion rates $k_{\mathrm{Nat,F}}$ and $k_{\mathrm{Nat,S}}$ to average lymphocyte lifetimes $\tau_{\mathrm{F}} = 1/k_{\mathrm{Nat,F}}$ and $\tau_{S} = 1/k_{\mathrm{Nat,S}}$, respectively, one obtains $\tau_{\mathrm{F}} = 5 \pm 4$ wk and $\tau_{\mathrm{S}} = 6 \pm 4$ y, reflecting two clearly distinct lymphocyte classes with markedly different turnover timescales. The corresponding weighted mean values of $\gamma$, $k_{\mathrm{Nat,F}}$, and $k_{\mathrm{Nat,S}}$ were then fixed as constants for all subsequent analyses, based on the assumption of biological universality across different patient populations and datasets, leaving only the $k_{\mathrm{Rad}}$ as a free model parameter.

In the second step, fitting was performed on the full subgroup of the 56 filtered datasets that contained at least two data points collected during treatment. This allowed the estimation of individual $k_{\mathrm{Rad}}$ values for each dataset and facilitated their interpretation with respect to tumor site, field size, and treatment modality. The distribution of $k_{\mathrm{Rad}}$ is shown in Fig. 1d. An exemplary fit of the data from Cho et al. [41] is presented in Fig. 1e, demonstrating the comparison between the full four-parameter fit and the simplified one-parameter fit with only $k_{\mathrm{Rad}}$ as a free parameter. The fitted $k_{\mathrm{Rad}}$ values with corresponding uncertainties and individual model fits are given in the supplementary material, Figures S3 and S4, respectively. A schematic overview of the two-step fitting procedure is provided in Fig. S5 of the supplementary material.

*Validation of Model Predictive Performance*

To assess the predictive capability of the biokinetic RIL model at the individual-patient level, a leave-one-out validation was performed using a subset of ten patient-specific ALC datasets from Ebrahimi et al. (2021) [19]. For each patient, the parameter $k_{\mathrm{Rad}}$ was replaced by the median value derived from all remaining cases, and the model was then used to simulate the lymphocyte dynamics of the omitted individual without further fitting. Model accuracy was evaluated by comparing the predicted and observed ALC at the last available measurement point for each patient, using mean absolute error (MAE) as a performance metric because of its robustness to outliers.

## Results

### *Radiation damage to lymphocytes is irradiation site specific*

To examine variations in radiation-induced lymphocyte depletion across different target locations, the datasets of patients ALCs were grouped based on the irradiated organ, and the median $k_{\mathrm{Rad}}$ for each organ was calculated. Table S2 summarizes these results. Datasets with data on multiple irradiated sites [36,40] were excluded from consideration.

Among the analyzed groups, brain and breast irradiation were associated with the lowest median $k_{\mathrm{Rad}}$ values ($0.12 - 0.14$ wk$^{-1}$), indicating relatively modest radiation-induced lymphocyte depletion. In contrast, irradiation of the esophagus, lungs, liver, and head & neck regions resulted in the highest $k_{\mathrm{Rad}}$ values ($0.40 - 0.46$ wk$^{-1}$), corresponding to more pronounced lymphocyte depletion. These sites also exhibited the lowest model-predicted EoT-RLC (approximating the ALC nadir), with median values around $0.16 - 0.26$, compared to over $0.5$ in brain and breast cases.

The relationship between the model-predicted EoT-RLC and the organ-specific median $k_{\mathrm{Rad}}$ values are illustrated in Figure 2a. The observed data from the fits to individual cohorts align closely with the trajectory predicted by the biokinetic model under the assumption of a fixed treatment duration of $\mathrm{T} = 5.7$ wk, which corresponds to the median therapy course length across all analyzed datasets. Results for individual datasets, highlighting variability within each organ group, are provided in the supplementary material, Figure S6.

*Response to total body irradiation reveals lymphocyte radiosensitivity*

Ellsworth et al. [40] provided an ALC dataset for patients treated with TBI, where all circulating lymphocytes receive a uniform, large radiation dose. This unique condition allows a lymphocyte radiosensitivity estimate based on observed depletion patterns. Given the specific fractionation schedule used (2 fractions per day over four days) it was necessary to adapt the model accordingly. The time axis of the dataset was converted from weeks to fractions, assuming 14 fractions per week, rescaling both data points and physiological parameters $k_{\mathrm{Nat,S}}$ and $k_{\mathrm{Nat,F}}$ accordingly. Applying the adapted model yielded an estimated per-fraction radiation-induced lymphocyte depletion rate of $k_{\mathrm{Rad}}^{\mathrm{Fx}} = 0.56 \pm 0.04$, which is a dimensionless quantity.

Assuming a linear survival model, the fractional depletion of lymphocytes per fraction is related to the delivered fraction dose $d = 1.5$ Gy and radiosensitivity parameter α through the equation

$$k_{\mathrm{Rad}}^{\mathrm{Fx}} = 1 - e^{-\alpha d} \qquad (4)$$

Using the estimated value $k_{\mathrm{Rad}}^{\mathrm{Fx}}$, the radiosensitivity was calculated as $\alpha = 0.55\ \pm 0.06$ Gy$^{-1}$. Alternatively, fitting the data directly with the linear survival model gave a lower estimate of $\alpha_{\mathrm{linear}} = 0.330\ \pm 0.015$ Gy$^{-1}$. The model fit to the TBI data is shown in Figure 2b.

*Model validation*

To evaluate predictive accuracy on the individual level, a leave-one-out cross-validation was performed (see *Materials and Methods*) on a cohort of ten esophageal cancer patients treated with proton therapy, as reported by Ebrahimi et al. (2021) [19]. Figure 3a shows the comparison between predicted and observed ALC values at the final measurement point. The model correctly reproduced RIL grade in 7 out of 10 patients, with a MAE of $0.12$ cells/nl. This demonstrates the model's ability to quantitatively predict EoT-ALC, RIL severity, and the lymphocyte-sparing effect of particle therapy on an individual basis.

*Particle therapy results in reduced lymphocyte depletion*

Eight publications [17,18,20–23,42,43] provide a direct comparison of ALC dynamics between patients treated with photon therapy and those treated with particle therapy (protons or carbon ions). Figure 3b presents the comparison of fitted $k_{\mathrm{Rad}}$ values between photon- and particle-treated patient groups for each of the referenced studies.

On average, particle therapy was associated with significantly lower lymphocyte depletion rates: the weighted mean of $k_{\mathrm{Rad}}$ across particle therapy datasets was $0.41 \pm 0.03$ wk$^{-1}$, compared to $0.57 \pm 0.03$ wk$^{-1}$ for photon therapy. The individual ratios $k_{\mathrm{Rad}}^{\mathrm{particles}} / k_{\mathrm{Rad}}^{\mathrm{photons}}$ from each study ranged from 0.62 to 0.90 across studies, with a weighted mean of 0.69 ± 0.06. This indicates, on average, a lymphocyte depletion per unit time about 30% lower in particle therapy compared to photon-based treatment.

*Early phase ALC allows to select patients benefiting from particle therapy*

To illustrate the model's predictive utility and lymphocyte sparing effect of particle RT on the individual level, we analyzed a cohort of ten esophageal cancer patients treated with proton therapy from Ebrahimi et al. (2021) [19]. For each patient, $k_{\mathrm{Rad}}$ and EoT-ALC were estimated from model fits.

Figure 3c presents the relationship between model predicted values of baseline ALC and EoT-ALC, together with the predicted EoT-ALC values assuming the same patients had been treated with photons—simulated by increasing $k_{\mathrm{Rad}}$ by $1/0.7 \approx 43\%$, consistent with the observed ~30% lymphocyte-sparing effect of particle therapy. The model predicts that one patient with grade (G) 2 RIL would develop G3 if treated with photons instead of protons, and that five of the seven G3 cases would escalate to grade 4 under photon therapy. The model-predicted dependencies of EoT-ALC on baseline ALC, assuming $k_{\mathrm{Rad}} = 0.46$ wk$^{-1}$ (median of the ten cases) for protons and $k_{\mathrm{Rad}} = 0.66$ wk$^{-1}$ for photons, closely follow the trend of individual patient data.

To further visualize the predicted sparing effect, Figure 4a presents a model-based comparison of ALC values during radiotherapy under particle versus photon therapy.

The horizontal axis represents ALC predicted for photon therapy (with $k_{\mathrm{Rad}} = 0.46$ wk⁻¹, the median for esophageal cancer), while the vertical axis shows ALC predicted for particle therapy (with reduced $k_{\mathrm{Rad}}^{\mathrm{particles}} = 0.7 \times k_{\mathrm{Rad}}^{\mathrm{photons}}$). Two representative baseline ALC values were simulated: $L_0 = 1.6$ cells/nl (typical for esophageal cancer, blue curve) and $L_0 = 1.0$ cells/nl (corresponding to pre-existing lymphopenia, red curve). Both curves lie consistently above the line of identity (red dashed), indicating systematically higher ALC values during treatment with particles. The blue curve demonstrates that for patients with normal baseline lymphocyte counts, ion therapy reduces the severity of lymphopenia by approximately one grade. In patients with already low baseline ALC, the sparing effect is smaller, yet even in this case, particle therapy prevents progression to G4 RIL.

To investigate the implications of such sparing in terms of lymphopenia induction, we combined these findings with an analysis of the variability of baseline ALC within patient cohorts. Using esophageal cancer as an example, we adopted $k_{\mathrm{Rad}} = 0.46$ wk⁻¹ from the model fit and $L_0 = 1.6$ cells/nl from inspection of pre-therapy ALC data [16], representing the respective median values across available esophageal cancer datasets. The model was then applied to map the baseline ALC distribution before therapy to post-RT ALC distributions for photons and particles (Fig. 4b). Here we assume that the baseline ALC distribution follows a gaussian shape with a mean of $L_0 = 1.6$ cells/nl and a dispersion of 30% of $L_0$, estimated from the variability of baseline ALC data across the GSI database. For particle therapy, $k_{\mathrm{Rad}}$ was assumed to be reduced by 30% (to 0.32 wk⁻¹), consistent with the previously established lymphocyte-sparing effect.

Figures 4c and 4d illustrate the resulting stratification of the baseline ALC distribution by post-RT lymphopenia grades. For photons, 15% and 84% of patients are predicted to develop grade 4 and grade 3 lymphopenia, respectively, whereas for particles these fractions decrease to 3% and 65%. Thus, 77% of photon-treated G4 patients would shift to G3 with particles, and 37% of G3 patients would shift to G2. Overall, particle therapy is predicted to reduce lymphopenia severity by one grade in ~50% of esophageal cancer patients.

These estimates are based on median values of $L_0$, $k_{\mathrm{Rad}}$ and therapy duration for esophageal cancer data and serve as illustrative examples. The parameters vary considerably within and across tumor sites and patient cohorts, leading to variability in predicted lymphopenia grades (Table S3 in the supplementary material). Accurate probabilities of RIL grade improvement therefore require applying the model to case-specific parameter values.

*Subgroup-specific analysis of lymphocyte depletion rates*

Subgroup-specific analyses were performed using datasets stratified by clinical or biological characteristics other than cancer site or treatment modality. The corresponding dataset selection is outlined in the supplementary material, Table S4. Briefly, subgroup comparisons revealed that baseline ALC consistently influenced depletion rates, whereas fractionation scheme (hypofractionation vs. conventional) showed no measurable impact. Given the limited number of available subgroup datasets, these findings should be interpreted cautiously and regarded as indicative rather than conclusive.

## Discussion

*Physiological parameter estimations*

The estimated parameters provide insight into the physiological turnover of circulating lymphocytes and their recovery dynamics after radiation exposure. An important question is how these inferred lymphocyte classes correspond to known immunological cell populations. The two-component approach effectively captures the well-established fact that relevant timescales of lymphocyte turnover can vary by at least an order of magnitude among different classes. In this way, the model effectively reflects the kinetic behavior rather than aiming for a one-to-one correspondence with specific immunological subtypes. Natural killer (NK) cells are typically short-lived, with turnover times around 2 weeks [44–46]. Naive B and T cells have lifespans ranging from months to several years [47–51]. Memory T cells show variable persistence: studies report half-lives from 30 to 160 days for short-lived memory cells [49,52], while long-lived memory T cells can persist for 8–15 years [53]. Boggen [54] identified two T cell subpopulations with distinctly different lifespans, estimating mean lifespans of 1.1 years and 6.3 years. Similar diversity is seen in B cell populations, with reconstitution timelines post-transplantation ranging from several months to over two years [48,51]. Based on these data, one may be tempted to associate FRL with NK cells and short-lived memory T cells, while SRL may correspond to naive and long-lived memory T and B cells. However, given the model's abstraction and the variability in reported values, we refrain from making definitive immunophenotypic assignments, and consider the parameterization as an effective approach.

*Cancer site dependence of lymphocyte depletion*

In our analysis, a clear dependency emerged between the irradiated anatomical region and the parameter $k_{\mathrm{Rad}}$ , which in turn influenced the model-predicted EoT-RLC.

Cancer sites associated with low $k_{\mathrm{Rad}}$ values, such as brain and breast, also exhibited relatively high EoT-RLCs—suggesting milder lymphopenia both during and after therapy. Conversely, cancers of the pancreas, pelvis, head & neck, liver, lung, and esophagus were associated with progressively higher $k_{\mathrm{Rad}}$ values and correspondingly lower EoT-RLCs, reflecting more pronounced lymphocyte depletion. This inverse correlation between $k_{\mathrm{Rad}}$ and EoT-RLC (c.f. Fig. S4) highlights a non-ambiguous link of the lymphocyte depletion rate during therapy with its cumulative impact by treatment end.

This variability in $k_{\mathrm{Rad}}$ across cancer sites underscores the relevance of treatment location in determining systemic immunological impact. Contributing factors likely include differences in typical radiation field sizes, prescribed doses, and the proximity of lymphocyte-rich tissues (e.g., bone marrow, spleen, lymph nodes), as well as regional blood flow and vascularization. A more detailed qualitative discussion of these determinants is provided in previous studies [16,19,55].

*Lymphocytes radiosensitivity estimation*

From the fitting of the TBI dataset, the radiosensitivity parameter $\alpha$ of the linear cell survival model was estimated as $\alpha = 0.55 \pm 0.06$ Gy$^{-1}$. This result is consistent with values reported in the literature [10,56]. An extended comparison of this estimate with experimental data is provided in the supplementary material, Figure S7.

Notably, the $\alpha$ value predicted by the biokinetic model is significantly higher than that obtained from a direct fit using the linear survival model alone ($\alpha_{\text{linear}} = 0.330 \pm 0.015$ $Gy^{-1}$). The inclusion of recovery mechanisms in the biokinetic model significantly shifts the estimated $\alpha$ value. Notably, both models show systematic discrepancies in describing the data, likely indicating that these models are oversimplified and do not capture additional dynamical processes.

*Lymphocyte sparing in particle therapy*

The type of radiation used in therapy can significantly influence the incidence and severity of RIL. Proton and carbon ion therapies are expected to be more favorable for lymphocyte preservation due to their reduced irradiated volume and lower integral body dose required to deliver an equivalent biological dose to the PTV. These dosimetric advantages translate into less exposure of circulating lymphocytes and lymphocyte-rich organs, aligning with clinical observations of reduced RIL in patients treated with particle therapy.

The observed differences in ALC dynamics between photon and particle treatments are predicted to be clinically meaningful. Particle therapy was associated with a significantly reduced lymphocyte depletion rate, resulting in a systematic shift of post-RT ALC distributions toward higher values. Consequently, the probability of developing G4 lymphopenia was substantially decreased, and a considerable proportion of patients experienced a one-grade reduction in RIL severity. Such improvements in ALC are expected to lower the overall incidence of severe lymphopenia (G3–G4) and to enhance patient resilience to immunotherapy.

The sparing of immune cells in particle therapy may enhance the efficacy of combined immunotherapy and radiotherapy by preserving T cell populations, particularly in lymphocyte-rich regions such as tumor-draining lymph nodes. This could promote sustained systemic immune responses and improve tumor control, including the potential for abscopal effects [57], expanding the conditions under which a systemic immune response can be achieved, enabling more adaptable treatment strategies and benefiting a broader patient population [18,58].

The presented model can be useful in clinical decisions on whether patients should be selected for particle therapy or not. Based on clinical experience, for patient selection criteria the therapy specific depletion rate $k_{\mathrm{Rad}}$ can be determined. For further patients, based on their baseline ALC $L_0$ or data gained in an early phase of therapy, a personalized prediction of EoT-ALC for both photon and particle therapy can be predicted. In particular patients who are projected to receive mild instead of severe lymphopenia when using particles, this may be a warranted clinical decision.

*Limitations of the Biokinetic RIL model*

While the presented biokinetic model effectively captures key trends in lymphocyte count dynamics during and after radiotherapy, several simplifications were made to maintain analytical tractability and avoid overfitting, especially given the limited size and resolution of available datasets.

First, the model groups lymphocytes into only two broad classes—fast-recovering and slow-recovering—despite the well-established heterogeneity of lymphocyte subtypes, including T cells, B cells, NK cells, and further subdivisions such as naive, memory, effector, and helper cells. This simplification abstracts the complex immunological landscape into kinetically distinct populations. Additionally, the model assumes rapid and complete redistribution of lymphocytes between the blood and lymphoid tissues, neglecting possible delays or retention effects in specific compartments. Radiation-induced lymphocyte depletion is modeled as a continuous and uniform process, which does not reflect the fractionated nature of clinical radiotherapy or the known delay between radiation exposure and subsequent cell death. Furthermore, lymphocyte production is assumed to occur at a constant rate, omitting potential radiation-induced suppression or feedback-driven stimulation of lymphopoiesis. The model also uses a single radiosensitivity parameter $k_{\mathrm{Rad}}$ for both lymphocyte classes, although radiosensitivity is known to vary across different lymphocyte populations. In addition, several methodological simplifications were introduced to facilitate robust parameter estimation. The natural death rates $k_{\mathrm{Nat}}$ and the baseline classes fraction $\gamma$ were fixed across all datasets, limiting the model's flexibility to capture inter-individual variability. ALC values were converted to RLC by normalizing all data points to the baseline ALC. This approach implicitly assumes that the baseline measurement is accurate, which may introduce a systematic shift in RLC values if the initial value is imprecise. Finally, the parameter $k_{\mathrm{Rad}}$ remains a derived, abstract quantity that is not directly linked to physical treatment parameters such as dose distribution, irradiated volume, or organ-specific exposure, limiting its interpretability in clinical or dosimetric terms. A more detailed discussion of the model's limitations, along with justifications and potential model extensions is provided in supplementary material, Table S5.

## Conclusion

The presented biokinetic model successfully reproduced the observed patterns of lymphocyte depletion and recovery and suggested that particle therapy offers a significant lymphocyte-sparing advantage over photon therapy, potentially improving immunotherapy efficacy and expanding the clinical applicability of combined treatment strategies. Additionally, the model enabled the estimation of lymphocyte radiosensitivity from data on TBI patients. Despite its simplifications, the model provides a robust and interpretable framework for quantifying immune suppression in radiotherapy, with potential applications in treatment planning and RIL severity prediction. By linking treatment modality to predicted lymphocyte preservation, it can be applied prospectively to assess whether particle therapy is likely to mitigate RIL in a given patient based on their baseline or early-treatment ALC measurements.

## Figures

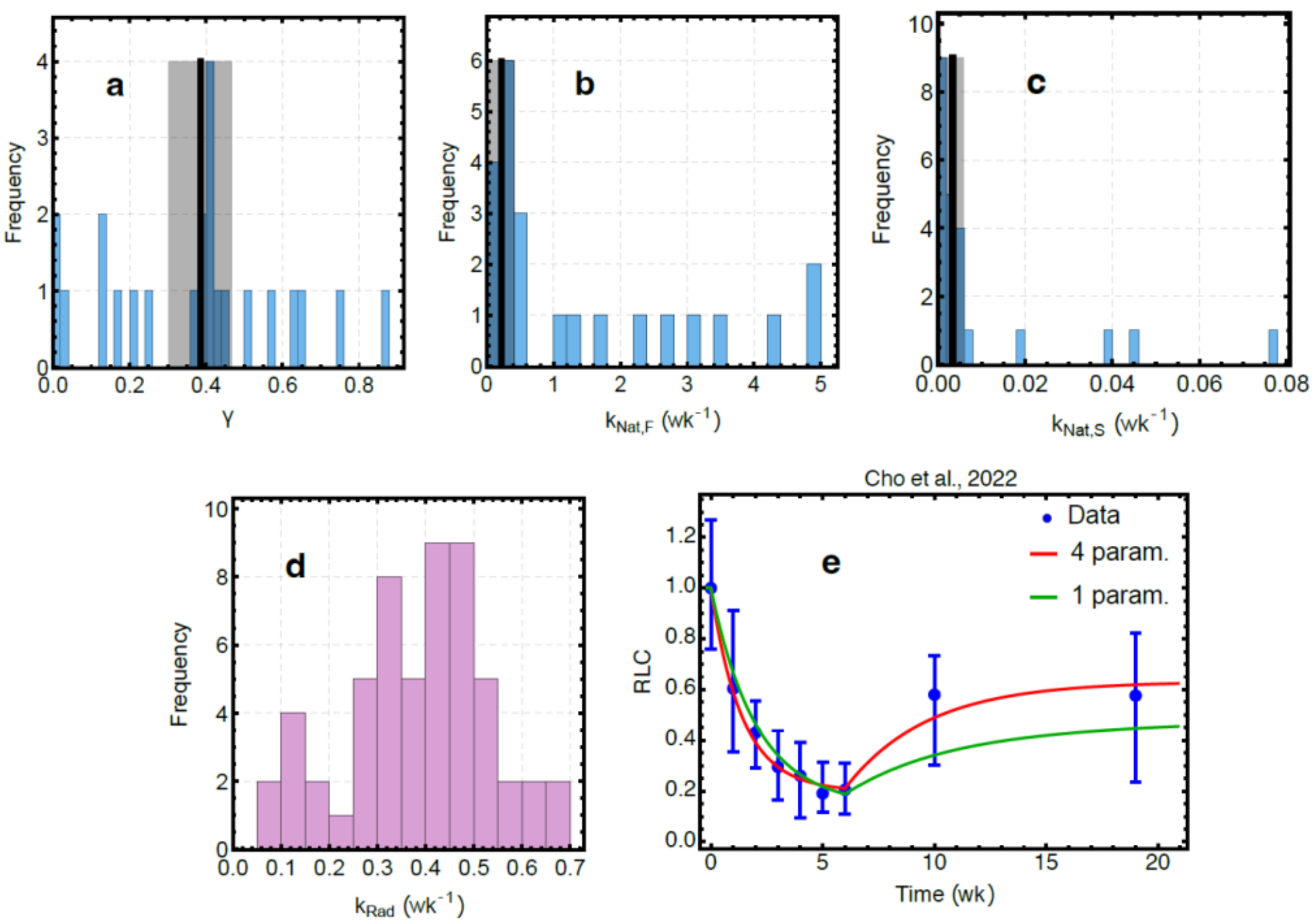

**Figure 1.** Distributions and representative fits of the estimated model parameters. (a–c) Results of the first step of the fitting procedure: distributions of (a) $\gamma$, (b) $k_{\mathrm{Nat,F}}$, and (c) $k_{\mathrm{Nat,S}}$. Weighted mean values are indicated by bold black vertical lines, and shaded grey areas represent uncertainty intervals. (d) Distribution of $k_{\mathrm{Rad}}$ estimated in the second step of the fitting procedure. (e) Dataset from Cho et al. [41] fitted with the full model including 4 free parameters and the reduced model with only $k_{\mathrm{Rad}}$ as a free parameter.

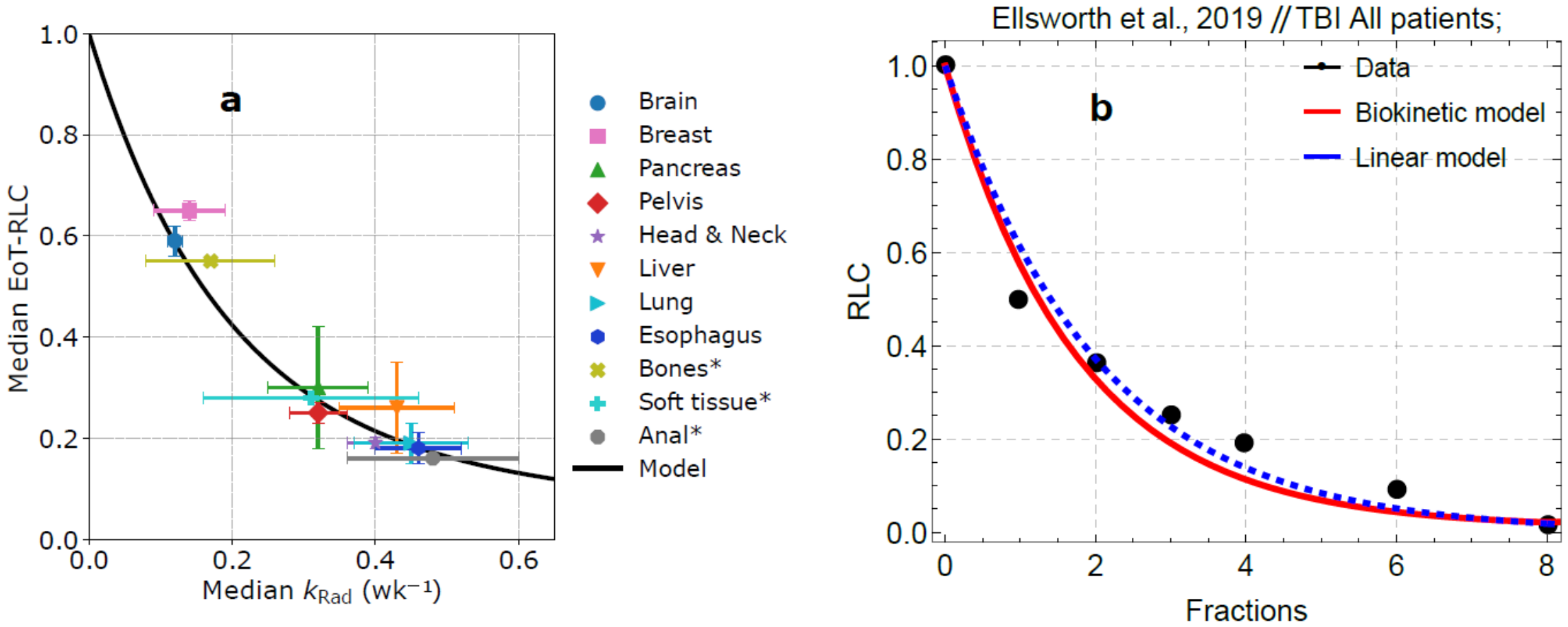

**Figure 2.** Applications of biokinetic model to clinical data. (a) Organ-specific relationship between median radiation-related lymphocyte depletion rate $k_{\mathrm{Rad}}$ and median model estimated EoT-RLC reflecting the acquired lymphopenia level. Points represent site-specific medians with error bars denoting the median absolute deviation (MAD). Cancer sites represented by a single dataset are marked with an asterisk in the legend. The solid black curve shows the model prediction for a treatment duration of $\mathrm{T} = 5.7$ wk, corresponding to the median across all datasets. (b) Fit of the biokinetic model (red) and a linear survival model (blue dashed) to RLC data from Ellsworth et al. [40] for patients treated with TBI. The time axis is expressed in the number of fractions (two fractions per day, eight total).

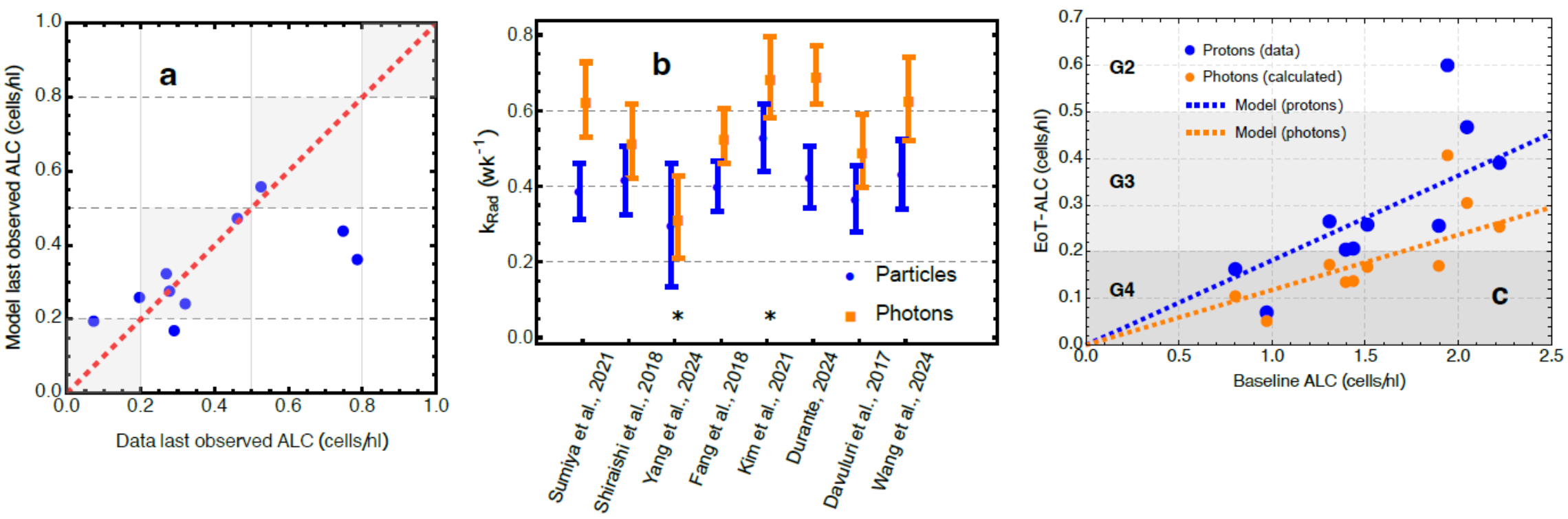

**Figure 3.** Model validation and comparison of lymphocyte depletion between photon and particle therapy. (a) Predicted versus observed ALC at the last available time point for a cohort of 10 esophageal cancer patients treated with protons (data from Ebrahimi et al., 2021 [19]). The red dashed line denotes identity. (b) Estimated $k_{\mathrm{Rad}}$ values for photon (orange) and particle (blue) therapy across published studies; error bars indicate fitting uncertainties. Asterisks denote datasets not referring to esophageal cancer. (c) Model-predicted EoT-ALC versus baseline ALC for 10 esophageal cancer patients treated with protons (data from Ebrahimi et al., 2021 [19]). Blue and orange dashed lines show model predictions for proton and photon therapy ($k_{\mathrm{Rad}}$ = 0.46 wk$^{-1}$ and 0.66 wk$^{-1}$, respectively). Shaded regions mark RIL grades G2–G4.

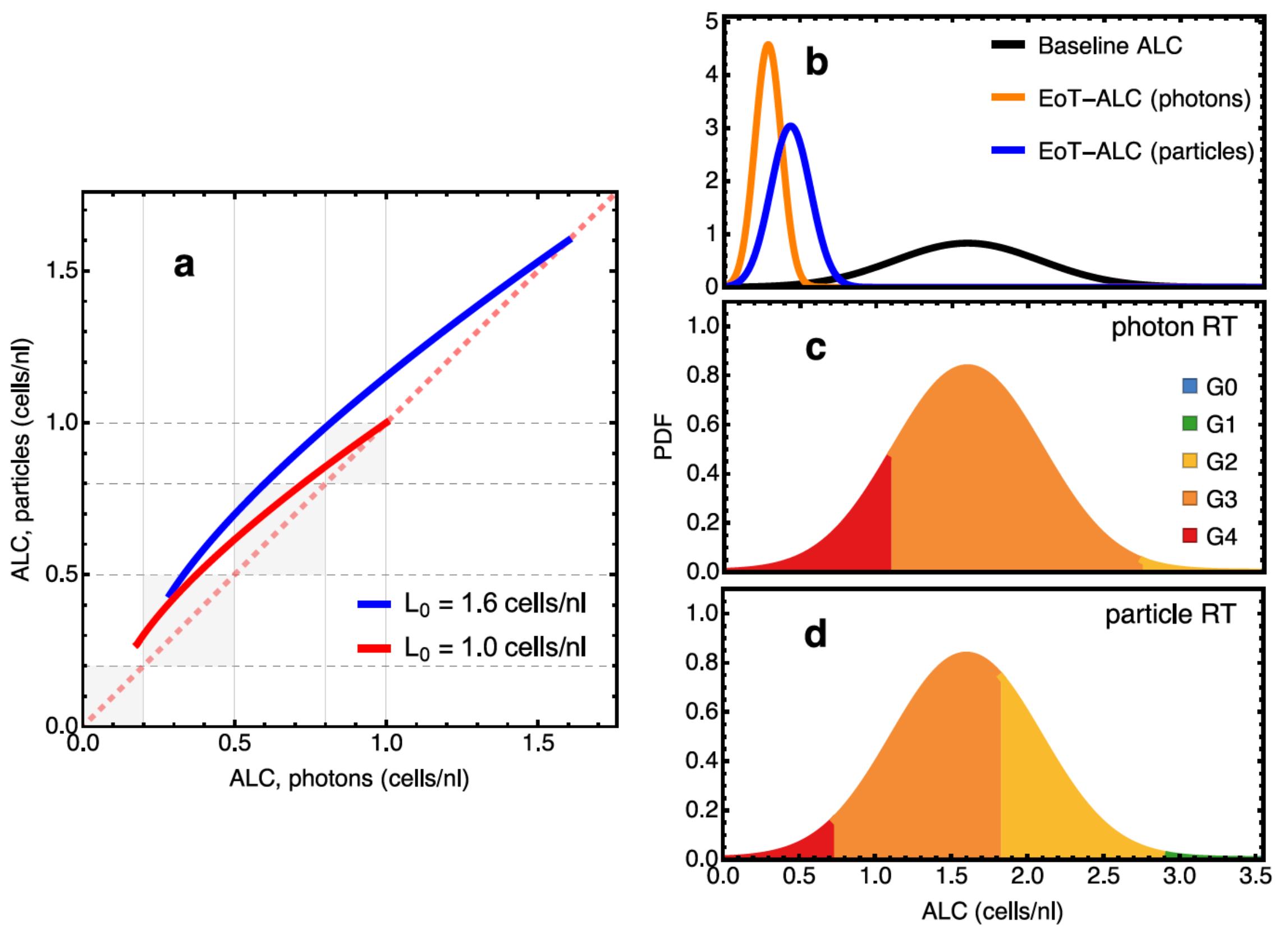

**Figure 4.** Model-based comparison of lymphocyte depletion dynamics between photon and particle RT. (a) Comparison of ALC predicted during RT under particle versus photon therapy. Two baseline ALC values were simulated: $L_0 = 1.6$ cells/nl (blue) and $L_0 = 1.0$ cells/nl (red). Curves above the identity line (dashed) indicate higher ALC during particle RT, typically corresponding to one-grade reduction in RIL severity. (b) Predicted probability density functions (PDFs) of ALC before (black) and after photon (orange) or ion (blue) therapy for esophageal cancer. (c,d) Mapping of baseline ALC distributions to post-RT lymphopenia grades (G0–G4) for photons and ions, respectively. Particle therapy results in a systematic right-shift of the distribution, reducing the fraction of G3–G4 lymphopenia cases.